\documentclass[aps,prd,twocolumn,superscriptaddress,showpacs,floatfix]{revtex4}

\usepackage{amssymb}
\usepackage{graphicx}
\usepackage{epsfig}
\newcommand{\ba}{\begin{eqnarray}}
\newcommand{\ea}{\end{eqnarray}}
\newcommand{\be}{\begin{equation}}
\newcommand{\ee}{\end{equation}}
\newcommand{\beq}{\begin{equation}} 
\newcommand{\eeq}{\end{equation}}   
\newcommand{\bea}{\begin{eqnarray}} 
\newcommand{\eea}{\end{eqnarray}}
\def\Li2{\hbox{Li}_2}

\begin{document}

\phantom{}

\vskip - 1cm

\hfill TTP09-11

\title{Strong and Electromagnetic $J/\psi$ and $\psi(2S)$ Decays into Pion and
Kaon Pairs}

\thanks{Work 
supported in part by BMBF grant 05HT6VKAI3,
EU 6th Framework Programme under contract MRTN-CT-2006-035482 
(FLAVIAnet)}


\author{Henryk Czy\.z}
\affiliation{Institute of Physics, University of Silesia,
PL-40007 Katowice, Poland.}
\author{Johann H. K\"uhn}
\affiliation{Institut f\"ur Theoretische Teilchenphysik,
Universit\"at Karlsruhe, D-76128 Karlsruhe, Germany.}


\date{\today}

\begin{abstract}
A combined analysis of the electromagnetic pion and kaon form factors in
the neighborhood of $J/\psi$ and $\psi(2S)$ and of the strong decay
amplitude of these resonances into kaons is presented. In the presence
of a large relative phase between strong and electromagnetic resonance
amplitudes the branching ratio, as measured in electron-positron
annihilation, receives an additional contribution from the interference
between resonance and continuum amplitude neglected in earlier papers.
Our study is model independent and does not rely on the SU(3) 
symmetry assumptions used in earlier papers.
We note that the large relative phase between strong and
electromagnetic amplitudes observed in earlier analyses is model
dependent and relies
critically on the specific assumptions on SU(3) symmetry and breaking.
\end{abstract}


\maketitle

\newcommand{\Eq}[1]{Eq.(\ref{#1})} 

\section{\label{sec1}Introduction}

Exclusive decays of $J/\psi$ and $\Psi(2S)$ into pseudoscalar meson
pairs have attracted considerable attention both from the theoretical
and the experimental side. In the case of $\pi^+\pi^-$ the branching
ratio was used to determine the pion form factor at fairly high
energies, which turned out to be significantly larger (see e.\ g. 
\cite{Milana:1993wk,Bruch}) than originally predicted in QCD. In the case of
charged and neutral kaons both strong and electromagnetic interactions
contribute with comparable strength and the branching ratio evidently
depends critically on their relative phase. In Refs.
\cite{Suzuki:1999nb,Rosner:1999zm,Seth:Jphi}
it has been argued this phase to be fairly large, close to 
$90^\circ$ or $270^\circ$ and interesting conclusions have been drawn for the
similar interplay between strong and weak phases in exclusive $B$ meson
decays.

Recently new experimental results for the branching ratios and new
measurements of the pion and  kaon production cross section close to
$\Psi(2S)$ became available. Furthermore there is the perspective of
improved measurements of these quantities both through the radiative
return \cite{Czyz:2000wh,Nowak,Czyz:2005as}
  at $B$ meson factories and from direct
scanning at BES. A detailed analysis of strong and electromagnetic
amplitudes which is less dependent on additional assumptions is
therefore appropriate.

In the following we will demonstrate that a previously neglected
interference term between continuum and resonance amplitude may
seriously affect the analysis of the kaon modes. Furthermore, the claim
of a large relative phase depends crucially on the assumption of an
extremely small $K_LK_S$ from factor, an assumption still to be
validated.

Our paper is organized as follows: In section II we present the
formalism used to describe both resonance decays and continuum cross
section in the neighborhood of the resonances.
 Section III is concerned with a
model-independent  extraction
of resonance parameters and form factors based on the most
recent data. Section IV contains a brief summary and our conclusions.

\section{\label{nr} Pseudoscalar pair production close to resonances}

\begin{figure}[h]
 \vspace{0.5 cm}
\begin{center}
\includegraphics[width=8.cm,height=8.cm]{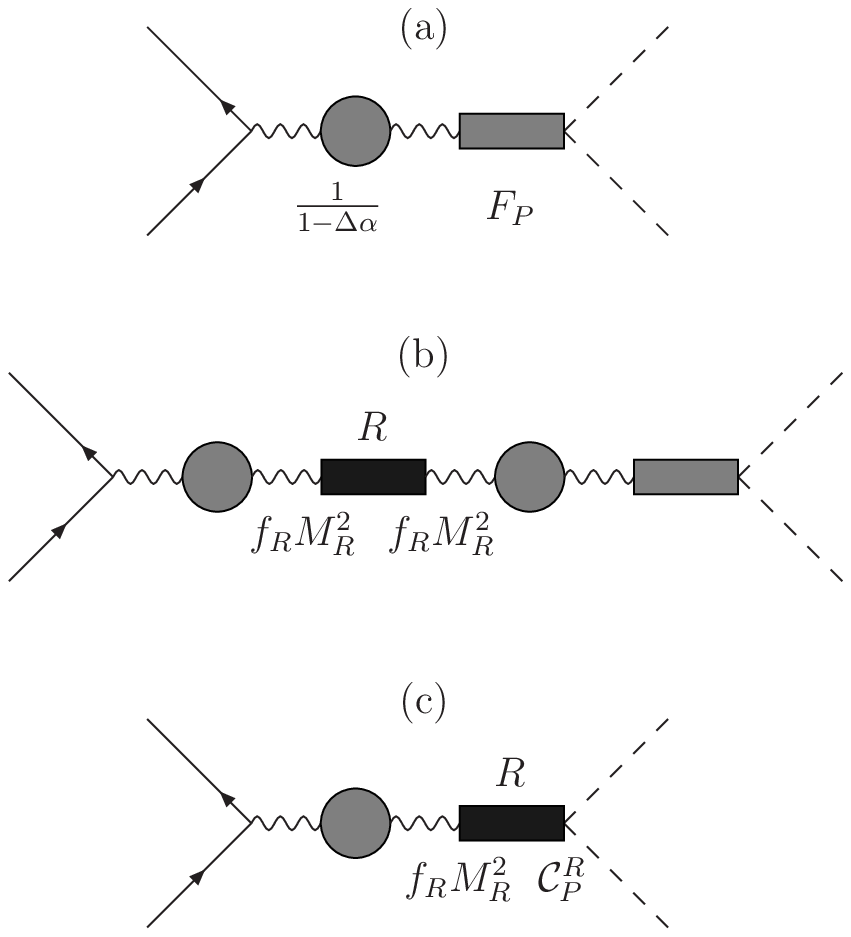}
\caption{(color online) The diagrams contributing to
 the cross section of the reaction
$e^+e^-\to P \bar P$. The grey bubble stands for the vacuum
 polarization, the grey box for the 'bare' form factor, the black
 box for the Breit-Wigner amplitude.\label{diagres}
}
\end{center}
\end{figure}

Various amplitudes can contribute to the production of
  hadronic final states
 close to a narrow resonance $R$: continuum production through the photon
 ($A_{QED}$, Fig. \ref{diagres}a), resonant production with
 electromagnetic decay ($A^R_{QED}$, Fig. \ref{diagres}b) and resonant 
 production with hadronic decay ($A^R_{QCD}$, Fig. \ref{diagres}c).
  All contributions involving virtual photons include 
 the vacuum polarization. The amplitudes 
 for pseudoscalar meson (denoted $P$) pair production, considered
 in this paper, (relative to the Born
 amplitude $\bar v \gamma_\mu u (q_1-q_2)^\mu/s$) are thus given by

  \bea
  A_{QED} &=& \frac{e^2}{1-\Delta\alpha} F_P \nonumber \\
  A^R_{QED}&=& \frac{e^2}{(1-\Delta\alpha)^2} 
    \frac{f_R^2 M_R^4}{s-M_R^2+i \Gamma_R M_R} \frac{1}{s} F_P \nonumber \\
 A^R_{QCD}&=& \frac{e}{1-\Delta\alpha} \ \ 
  \frac{f_R M_R^2\ {\cal C}_P^R}{s-M_R^2+i \Gamma_R M_R} \ .
\label{amplitudes}
  \eea

The $s$ dependence of $\Delta\alpha \equiv {\rm Re} \Pi(s)$ and
 of the form factor $F_P$ is implicitly understood. 
 $M_R$ and $\Gamma_R$  denote
  mass and width of the narrow resonance. In the absence of 
 dispersive contributions the coupling $f_R$ 
 of the virtual photon to the resonance
  is real, while the electromagnetic 
  form factor $F_P$ and the coupling $ {\cal C}_P^R$
 are complex. The contribution from the (lepton- and hadron-induced)
 vacuum polarization  is displayed explicitly, hence $F_P$ denotes
  the ``bare'' form factor. (Note that in \cite{Bruch} the ``dressed'' 
  form factor was used.) For the subsequent discussion it is convenient
to express the combination
  $f_R^2M_R^4/[s(1-\Delta\alpha)^2]$ by $3\sqrt{s}\ \Gamma_e^R/\alpha$
  and $ {\cal C}_P^R$ by $e  F_P f_R M_R^2 \ c_{P}^{R} /[s(1-\Delta\alpha)]$, where
  $\Gamma_e^R$ is the electronic
 width of the narrow resonance, and $c_{P}^{R} = A^R_{QCD}/A^R_{QED}$.

 The cross section for the reaction
$e^+e^-\to P \bar P$ can then be written in the form

  \bea
 &&\kern-20pt \sigma\left(e^+e^-\to P \bar P\right) =
\nonumber \\
 &&\frac{\pi\alpha^2}{3s}|F_P|^2 \ \beta^3\nonumber \\
 &&\times |\frac{1}{1-\Delta\alpha}+\sum_{R}\frac{3\sqrt{s}}{\alpha}
 \frac{\Gamma_e^R (1+c_P^R)}{s-M_R^2+i \Gamma_R M_R}|^2 = \nonumber \\
 \nonumber \\
  &&\frac{\pi\alpha^2}{3s}|F_P|^2 \ \beta^3\nonumber \\
 &&\times\Biggl(\frac{1}{(1-\Delta\alpha)^2} + \sum_{R}\Bigl\{\frac{9s}{\alpha^2}\frac{(\Gamma_e^R)^2}
 {(s-M_R^2)^2+\Gamma_R^2 M_R^2}\nonumber \\
 &&\kern+20pt\times \Bigl[|1+c_P^R|^2 + \frac{2\alpha M_R}
{3 \sqrt{s}(1-\Delta\alpha)}
 \frac{\Gamma_R}{\Gamma_e^R} {\rm Im}(c_P^R)\Bigr]\nonumber \\
&&\kern+20pt+ \frac{6\sqrt{s}\Gamma_e^R}{\alpha(1-\Delta\alpha)}
 \frac{\left(1+{\rm Re}(c_P^R)\right) \left(s-M_R^2\right) }
 {(s-M_R^2)^2+\Gamma_R^2 M_R^2}\Bigr\}\Biggr)
 \ ,\nonumber \\ 
\label{cross}
   \eea
where $\beta = \sqrt{1-\frac{4m_P^2}{s}}$. For muons the factor
 $\beta^3$ is replaced by  $4\beta(1+2m_\mu^2/s)$.
 The  narrow resonances are well separated, thus in 
 Eq.(\ref{cross}) their interferences are neglected.
 We also neglect here small contributions from the imaginary part of 
 $\Delta\alpha$.

In the present discussion  the narrow resonances  $J/\psi$ 
 and $\psi(2S)$ are considered. 
 For pion pairs the direct couplings  $c_P^R$ of the narrow resonances
  to the hadronic final states 
 vanishes  as a consequence of isospin conservation.
 For kaons, however, the direct coupling is important
  and the
 relative magnitude and the phase of QED versus hadronic amplitude
 can only be obtained from a study combining measurements on and off resonance.
 Such an investigation might, furthermore, even allow to disentangle
 the $I=0$ and $I=1$ amplitudes that contribute to the electromagnetic
  form factor. For the muon the form factor is set to $1$. 
  For baryons, which are not studied in this paper,
 one receives contributions from the electric and magnetic form factors
 (see e.g. \cite{Nowak} and references therein). Also for baryons there exist
 a direct coupling to the narrow resonances and both isospin zero 
 and isospin one contribute to the form factors, hence the discussion
 is quite similar to the case of kaons.

 The interference term between continuum and any of the resonances
  consists of two parts: one  proportional to ${\rm Im}(c_P^R)$
  and the second  one proportional to $\left(1+{\rm Re}(c_P^R)\right)$.
   The term  $\sim {\rm Im}(c_P^R)$
 contributes both to the integrated resonant cross section 
 and to its peak value. These quantities thus are 
 not, as it is often assumed (see e.g. \cite{Rosner:1999zm,Suzuki:1999nb,Seth:Jphi}),
 proportional
 to  $\Gamma(R\to P\bar P)$ which is defined without the interference with
 the continuum: 
   \bea
      \Gamma(R\to P\bar P) = \Gamma^{QED} \times |1+c_P^R|^2 \ ,
   \label{width}
   \eea
where $\Gamma^{QED}$ represents the decay rate induced through
 $A^R_{QED}$ alone.
  The correct formula,
 to be used to calculate
  the branching ratios of $J/\psi$ and $\psi(2S)$ decays,
  which are extracted from the integrated cross section,
  thus reads
   \bea
      \Gamma(R\to P\bar P) &=& \Gamma^{QED} \times \Bigl[|1+c_P^R|^2
\nonumber \\
  &&+ \frac{2\alpha M_R}{3 \sqrt{s}\left(1-\Delta\alpha\right)}
 \frac{\Gamma_R}{\Gamma_e^R} {\rm Im}(c_P^R)\Bigr] \ .
   \label{widthcorrect}
   \eea
The additional term
 correctly takes into account the interference effects between
 continuum
 and resonance. 

From Eq.(\ref{cross}) it is evident that electromagnetic and hadronic
 decay amplitudes, $A^R_{QED}$ and  $A^R_{QCD}$, can only be disentangled,
 if both the even part (contributing to the peak value, the branching
 ratio and the far off resonance behaviour) and the odd part 
 (in the neighborhood of the resonance) are measured.

 At present the complete 
 analysis  can be performed only for the $\psi(2S)$
 decaying to charged kaons
 where an off peak measurement
 of the cross section is available \cite{Pedlar:2005sj}.
 For the $\psi(2S)$ decay to neutral kaons as well as for the
 $J/\psi$ decays off peak data are not available
 and the only information about the coupling $c_P^R$ 
  comes from the branching ratios. We will return to this point in
 the next  Section.
  
 \section{\label{decays} Analysis of resonance parameters and form factors}

 In the following we shall present an analysis of the resonance 
 parameters valid for $J/\psi$ and $\psi(2S)$ decays into 
 $\pi^+\pi^-$, $K^+K^-$ and $K^0\bar K^0$.
 Masses, decay rates, branching ratios, form factors and values of
 $\Delta\alpha$ are listed in Table \ref{tab:res}.

\begin{table}[hb]
\begin{center}
\begin{tabular}{|c|c|c|}
\hline
   & $J/\psi$ & $\psi(2S)$  \\
\hline
 $M$ [MeV] & 3096.916$\pm$0.011  &  3686.09$\pm$0.04 \\
\hline
 $\Gamma_{ee}$ [keV] & 5.55$\pm$0.14$\pm$0.02   & 2.38$\pm$0.04 \\
\hline
 ${\cal B}(e^+e^-)$[\%] & 5.94$\pm$0.06  & 0.752$\pm$0.017\\
\hline
 ${\cal B}(K^+K^-)$[10$^{-5}$] & 23.7$\pm$3.1  & 6.3$\pm$0.7\\
\hline
 ${\cal B}(K^0_SK^0_L)$[10$^{-5}$] & 14.6$\pm$2.6  & 5.4$\pm$0.5\\
\hline
 ${\cal B}(\pi^+\pi^-)$[10$^{-5}$] & 14.7$\pm$2.3  & 0.9$\pm$0.5\cite{Dobbs2}\\
\hline
 $\sigma(K^+K^-)$[pb] & -  & 5.7$\pm$0.8 
 \cite{Pedlar:2005sj} \\
\hline
 $\sigma(K^0_SK^0_L)$[pb] & -  & $<$ 0.74 (90\% C.L.) 
 \cite{Dobbs:2006fj} \\
\hline
 $\sigma(\pi^+\pi^-)$[pb] & -  & 9.0$\pm$2.2 
 \cite{Pedlar:2005sj} \\
\hline
 $\Delta\alpha$ &  0.02117 \cite{Jeg_web,Czyz:2005as} &
     0.02219\cite{Jeg_web,Czyz:2005as}\\
\hline
\end{tabular}
\caption{{ Resonance parameters, vacuum polarization and cross sections used
  in this paper. If not stated differently the values are taken from
  \cite{Amsler:2008zz}. The cross sections are measured at $\sqrt{s} = 3.671$~GeV .}}.
\label{tab:res}
\end{center}
\end{table}
Let us first discus the decays to $\pi^+\pi^-$. Isospin symmetry forbids the 
 direct hadronic decay, whence $A^R_{QCD}=0$. 
 In this case the form factor can be directly derived from the branching
  ratios listed in Table \ref{tab:res}:

 \bea
  |F_\pi|^2 = \frac{4 {\cal B}(R\to \pi^+\pi^-)}{\beta^3_\pi 
 {\cal B}(R\to e^+e^-)}
 \ .
  \label{ffpsi1}
 \eea

 For some time it has been argued 
 \cite{Milana:1993wk,Bruch} that the result 
  $|F_\pi(\sqrt{s}=M_{J/\psi})|^2 = (10.0\pm 1.6)\cdot10^{-3}$,
 as derived from this consideration, is surprisingly large, 
  when confronted \cite{Bruch}
 with predictions based on the asymptotic pion wave function
 and derived in perturbative QCD. 
 However, an independent and direct measurement of 
 $\sigma(e^+e^-\to\pi^+\pi^-)$ in the 
 neighborhood of $\psi(2S)$ 
 \cite{Pedlar:2005sj} is consistent with this relatively large form factor.
 Thus we shall adopt $A^R_{QCD}(\pi^+\pi^-)=0$
   in this paper, 
 although an independent
 measurement of the pion form factor in the $J/\psi$ region would be highly
  desirable. The results for 
 $|F_\pi(\sqrt{s}=M_{J/\psi})|^2$ and 
 $|F_\pi(\sqrt{s}=3.671 {\rm GeV})|^2$ are listed in Table \ref{tab:res1}.
   Our value for the pion form factor 
  $|F_\pi|^2 = (5.92\pm 1.46)\cdot 10^{-3}$,
  as obtained from the off-resonance
 cross section measurement \cite{Pedlar:2005sj} and listed in 
 Table \ref{tab:res}, is
 about 6\% higher than the result quoted in \cite{Pedlar:2005sj}
 as a consequence of the interference between the resonance and
 the continuum neglected in \cite{Pedlar:2005sj}.

\begin{table}[hb]
\begin{center}
\begin{tabular}{|c|c|c|}
\hline
   & $J/\psi$ & $\psi(2S)$  \\
\hline
 $|F_\pi|^2$[10$^{-3}$] Eq.(\ref{ffpsi1}) & 10.0$\pm$1.6  
  & 4.8$\pm$2.73  \\
\hline
 $|F_\pi|^2$[10$^{-3}$] Eq.(\ref{cross}) & - 
  & 5.92$\pm$1.46 \\
\hline
 $|F_{K^+}|^2$[10$^{-3}$]Eq.(\ref{ffkp}) & -  & 4.0$\pm$1.3 
  \\
\hline
 $|F_{K^0}|^2$[10$^{-3}$]Eq.(\ref{modf0}) &  - &  $<$ 0.81 \\
\hline
\end{tabular}
\caption{{ Form factors obtained from our analysis.}}.
\label{tab:res1}
\end{center}
\end{table}

 Let us now move to the case of kaons. As stated in the Introduction,
 a completely independent determination of the three amplitudes
 $A_{QED}$,   $A^R_{QED}$ and  $A^R_{QCD}$, without further assumptions can
  only be obtained if both the even and the odd parts of the cross section
 are measured in the neighborhood of the resonance. In the first step
 we will discuss the
 decay $\psi(2S)\to K^+K^-$. The resonant symmetric part, which is determined
 by the branching ratio, receives a contribution from $|1+c_P^R|$
  and a contribution proportional to ${\rm Im}(c_P^R)$, which is often ignored.
 Its numerical relevance will be discussed below. For the ratio
 ${\cal B}(K^+K^-)/{\cal B}(e^+e^-)$ we find

  \bea
  R_+ \equiv \frac{4{\cal B}(K^+K^-)}{\beta^3_{K^+}{\cal B}(e^+e^-)}
  = |F_{K^+}|^2\left[ |1+ c_+|^2 
 + r {\rm Im} c_+   \right] \ ,
  \label{rkpl}
  \eea
 where for simplicity we use notation $c_+=c_{K^+}^{\psi(2S)}$. Moreover
 
  \bea
  &&\beta_{K^+}^3 = \left(1-\frac{4m^2_{K^+}}{M^2_{\psi(2S)}}\right)^{3/2}
  = 0.894
  \nonumber \\ && R_+ = (3.75\pm0.43)\cdot10^{-2}\nonumber \\
  &&r=\frac{2\alpha}{3(1-\Delta\alpha){\cal B}(e^+e^-)} =  0.663\pm 0.015 \ \ . 
\eea

Anticipating a value of $c_+$ with a modulus
  around 3 and a sizable imaginary part,
 it is clear that the interference term $ r {\rm Im} c_+$ cannot be neglected.

  In order to disentangle the form factor from the modulus and the phase
 of $|c_+|$, a measurement of $|F_{K^+}|$ far away from resonance and a second
 measurement closer to the peak would be required. This is illustrated
 by evaluating Eq.(\ref{cross}) off resonance, ($|E-M_R|\gg\Gamma_R$).
 In this case Eq.(\ref{cross}) can be cast into the form
 
 \bea
  S_+ - R_+\frac{\gamma^2 }{4} = |F_{K^+}|^2
  \left[1+\gamma(1+{\rm Re}c_+)\right]  ,
 \label{croffres}
 \eea
with 
\bea
 S_+ = \sigma(e^+e^-\to K^+K^-)(1-\Delta\alpha)^2/
 \left(\frac{\pi\alpha^2}{3s}\beta_{K^+}^3\right) 
\eea
and
\bea
\gamma = \frac{\Gamma_e}{E-M_R}\frac{3(1-\Delta\alpha)}{\alpha}.
\eea

 The cross section 
 $\sigma(e^+e^-\to K^+K^-)$, as defined  in Eq.(\ref{cross}),
 still contains all vacuum polarization corrections.
 When using experimental data one has to remember that it
 is often corrected for the leptonic part of the vacuum polarization.

 Given $|F_{K^+}|$ from one measurement far enough from the resonance
 to suppress the interference ($\gamma\to 0$),
  and performing another one with $E-M_R$ such that $\gamma\sim -1$
   Eq. (\ref{croffres}) will lead to a model-independent determination
  of $|F_{K^+}|^2$ and ${\rm Re}c_+$. Eq. (\ref{rkpl}) can then be used
 to determine ${\rm Im}c_+$ up to a twofold ambiguity.
  A measurement above the resonance would suffer from contributions from
  the radiative return to the resonance, thus it is easier to make the
  measurement below.  

\begin{figure}[ht]
 \vspace{0.5 cm}
\begin{center}
\includegraphics[width=8.cm,height=7.cm]{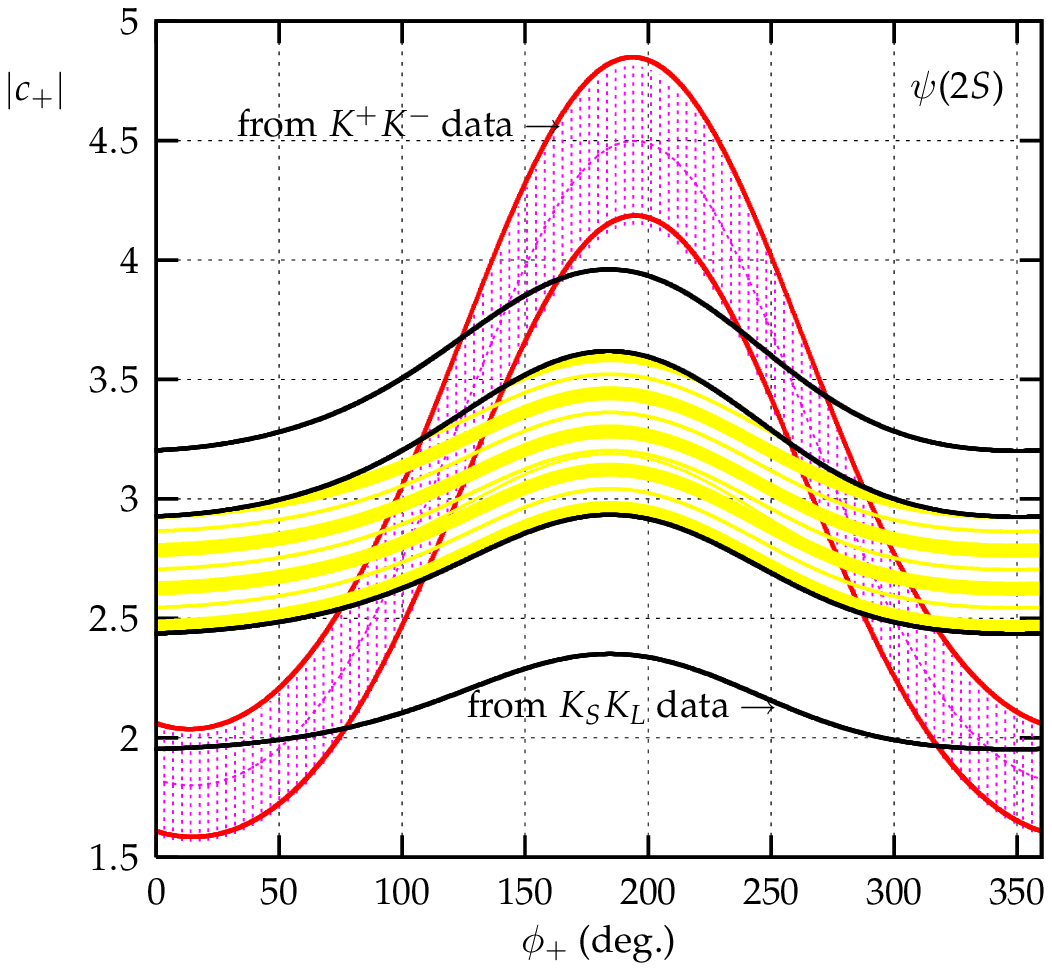}
\includegraphics[width=8.cm,height=7.cm]{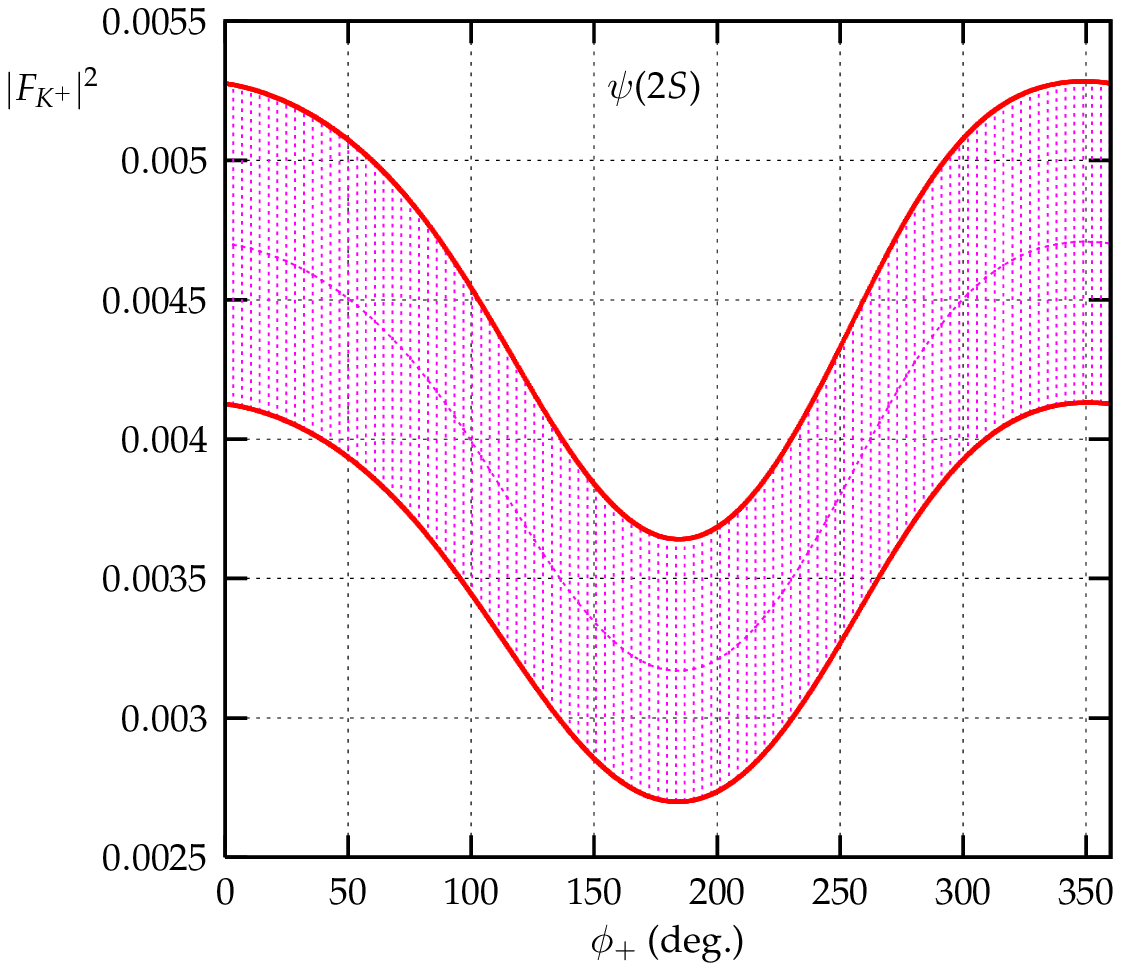}
\caption{(color online)
 The experimentally allowed regions of $|c_+|$, $|F_{K^+}|^2$ as functions
 of $\phi_+$.
 To obtain the vertically dashed regions 
  we used $K^+K^-$ data.
 The narrow horizontal band was obtained using $A_{QED}(K^0) =0$, while 
 the wider horizontal region was obtained using neutral kaon
 data and  in addition $|F_{K^+}|^2(\phi_+)$ (see text for details).
\label{fig:cphi}
}
\end{center}
 \vspace{0.5 cm}
\end{figure}

 The CLEO determination 
 \cite{Pedlar:2005sj} of the cross section 15~MeV below the $\psi(2S)$
 resonance leads to
  $\gamma = -0.063\pm 0.001$, $S_+ = (3.90\pm 0.52)\cdot10^{-3}$
  and $S_+ - R_+\gamma^2 /4 = (3.86\pm 0.52)\cdot10^{-3}$.
 In this case
  the $\gamma^2$ term is negligible and the term
  linear in $\gamma$ is a small correction. In contrast, for a value of
 $|E-M_R|$ say around 2 MeV and correspondingly $\gamma\sim -0.5$,
 the interference term is large and the measurement is sensitive to
  the product $\gamma{\rm Re}c_+$. Let us now evaluate
 $|c_+|$ and $|F_{K^+}|^2$ as a function of $\phi_+$ ($c_+=|c_+|e^{i\phi_+}$)
 for the present set of measurements. Combining 
 Eqs. (\ref{croffres}) and (\ref{rkpl}) one finds
 \bea
  &&\kern -40pt\frac{R_+}{S_+ - R_+\frac{\gamma^2 }{4}} = \nonumber \\ 
 &&\frac{ 1+ 2 |c_+| \cos(\phi_+) + |c_+|^2  + r |c_+|\sin(\phi_+)  }
     {1+\gamma(1+|c_+|\cos(\phi_+))} 
\label{modcphi}
  \eea
 and the values of $|c_+|$ can be determined as a function of its phase $\phi_+$
  (vertically dashed region in the upper plot of Fig.\ref{fig:cphi}). 

\begin{figure}[ht]
 \vspace{0.5 cm}
\begin{center}
\includegraphics[width=8.cm,height=7.cm]{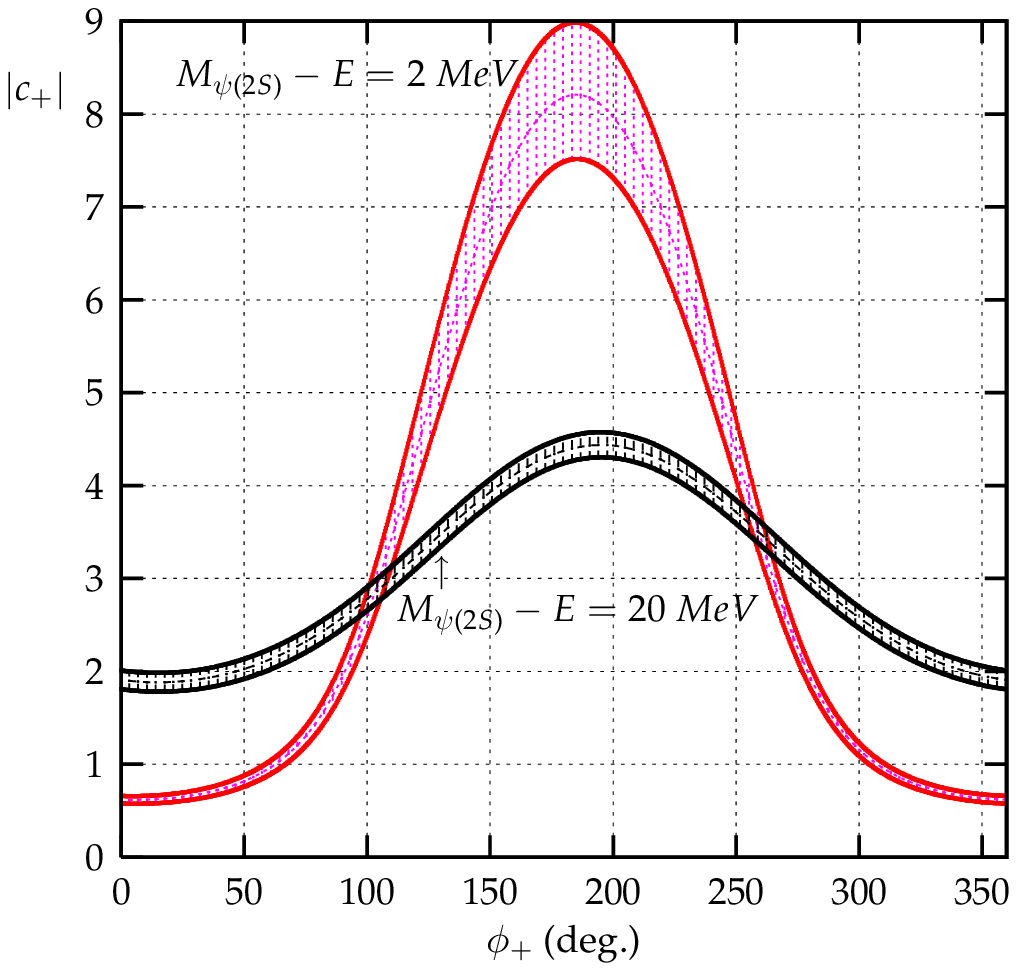}
\includegraphics[width=8.cm,height=7.cm]{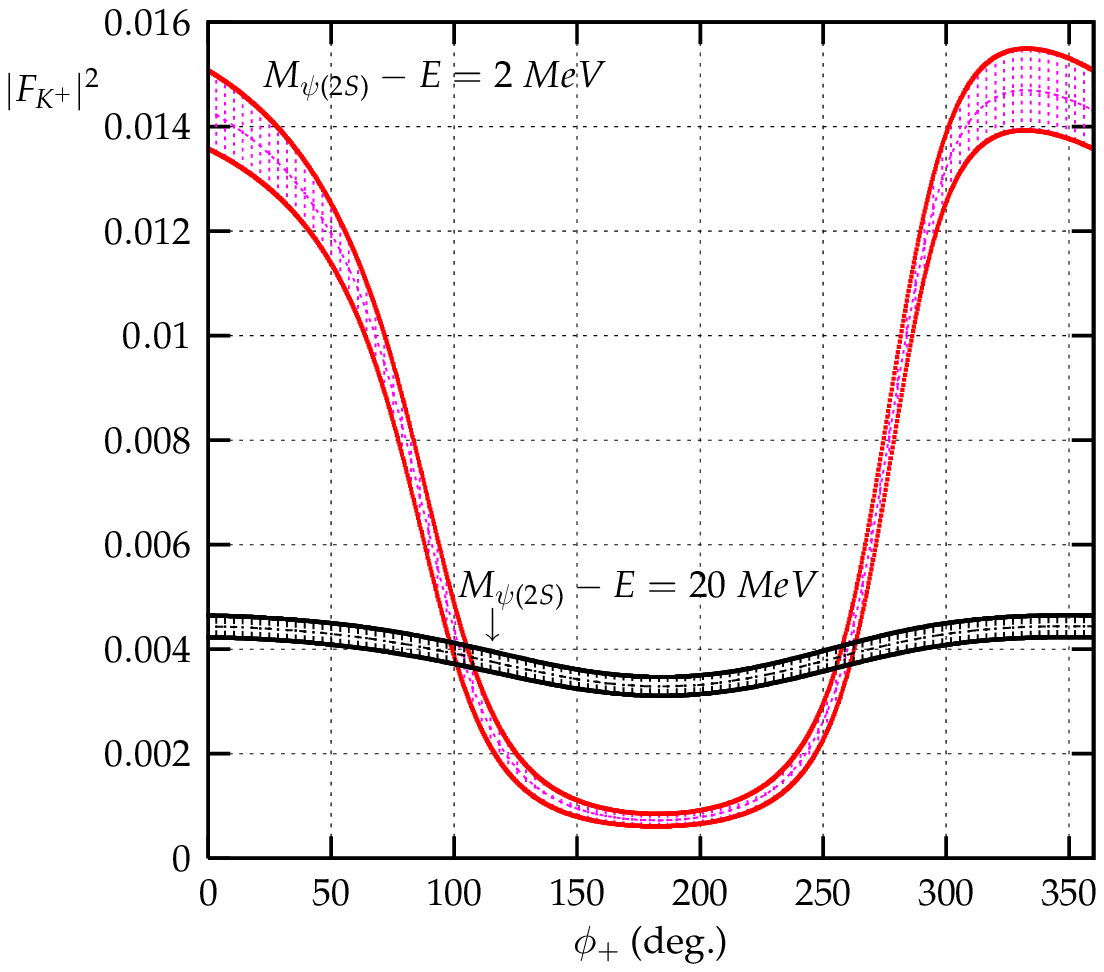}
\caption{(color online)
  Allowed regions of $|c_+|$, $|F_{K^+}|^2$ and $\phi_+$
 in the fictitious experimental situation described in the text.  
\label{fig:improv}
}
\end{center}
 \vspace{0.5 cm}
\end{figure}

As a consequence of the smallness of $\gamma$, the form factor $|F_{K^+}|^2$
 as extracted
 from Eq.(\ref{croffres}) is only moderately dependent on $\phi_+$ 
 (Fig.\ref{fig:cphi}). Nevertheless, the experimentally allowed 
 interference term
 might not be  negligible as assumed in \cite{Pedlar:2005sj}. Thus
 the error on $|F_{K^+}|^2$ as extracted from  $\sigma(e^+e^-\to K^+K^-)$
  has to be enlarged as compared to
  \cite{Pedlar:2005sj}. For an unknown phase $\phi_+$  the experimentally
 allowed region of $|F_{K^+}|$ is given by (see Fig. \ref{fig:cphi})

  \bea
   0.052 < |F_{K^+}| < 0.073
   \label{ffkp}
  \eea 
as compared to the value obtained in \cite{Pedlar:2005sj}

  \bea
   0.059 < |F_{K^+}| < 0.067 \ ,
  \eea 
which corresponds to the special case $\gamma(1+{\rm Re}c_+)=0$.
 As stated above, if a second measurement of $\sigma(e^+e^-\to K^+K^-)$
 closer to the $\psi(2S)$ would be available a model independent determination
 of ${\rm Re}c_+$, $|F_{K^+}|^2$ with a twofold solution
 for ${\rm Im}c_+$  would be feasible. This is illustrated
 in Fig. \ref{fig:improv} where the analysis based on two fictitious measurements
 \bea
 &&\kern-40pt\sigma(e^+e^-\kern-2pt\to\kern-2pt K^+K^-, E\kern-2pt=\kern-2ptM_{\psi(2S)}\kern-2pt-\kern-2pt 20 {\rm MeV}) \nonumber \\
 &&= (5.55\pm 0.28) pb  
 \eea
and 
 \bea
 &&\kern-40pt\sigma(e^+e^-\kern-2pt\to\kern-2pt K^+K^-, E\kern-2pt=\kern-2ptM_{\psi(2S)}\kern-2pt-\kern-2pt 2 {\rm MeV}) \nonumber \\
 &&= (7.68\pm 0.38) pb  
 \eea
 in combination 
with an improved determination of 
${\cal B}(\psi(2S)\to K^+K^-)=(6.3\pm0.35)\cdot10^{-5}$
 (we assume that the central remains unchanged and the error is
 reduced by a factor 2
  as compared to \cite{Amsler:2008zz}).  
  With a luminosity of  BES III \cite{Asner:2008nq} of
  $2.6\cdot10^3 {\rm pb}^{-1}/{\rm month}$ one expects about 1000 
 $K^+K^-$ events in 2 days of running, thus the proposed measurements
  are certainly possible. 
 As it is clear from  Fig. \ref{fig:improv} a remarkable accuracy
  for  $|c_+|$, $|F_{K^+}|^2$ and $\phi_+$ can be expected. Two solutions
  are allowed for $|c_+|$ and $\phi_+$ corresponding to two values of 
 ${\rm Im} c_+$.
 This ambiguity  can only be resolved
 with information coming from neutral kaon production.

For the moment, with only one off-resonance measurement available,
 one has to rely on the measurements of the $\psi(2S)$
 decay into $K^0\bar K^0$,
 limited information on the neutral kaon form factor $F_{K^0}$, 
 and  isospin symmetry relating the hadronic 
 amplitudes $A^R_{QCD}(K^+K^-)=A^R_{QCD}(K^0\bar K^0)$. The determination
 of $A^R_{QCD}(K^0\bar K^0)$ makes use of the upper limit 
  $\sigma(e^+e^-\to K^0_S K^0_L) < 0.74 {\rm pb}$
 at 90\% C.L.,  obtained \cite{Dobbs:2006fj}
 15~MeV below the resonance.
  In \cite{Dobbs:2006fj} this measurement was translated into
 the limit  $|F_{K^0}|^2 < 0.53\cdot10^{-3}$,
  neglecting the interference term in Eq.(\ref{cross}),
 i.e. by setting $\gamma=0$. Based on the smallness of  $|F_{K^0}|$
 one might then set $A^R_{QED}=0$ and derive the hadronic amplitude
 $A^R_{QCD}$ from the branching ratio into $K^0\bar K^0$.
 Using isospin invariance this would lead to 
   $|c_+F_{K^+}| = 0.179\pm 0.009$. The result for $|c_+|$ is shown
 in Fig. \ref{fig:cphi} as a narrow waved band.

  A more refined and conservative analysis, which is sensitive
 to the interference term and makes use only of the
 fore mentioned upper limit 
 on the cross section,
 leads in the first step 
 to 

  \bea
  \mid \frac{1}{c_0}\mid = \mid \frac{A^R_{QED}}{A^R_{QCD}} \mid < 0.187\pm0.008
 \ .
 \label{invc0}
  \eea

  The uncertainty comes from the error on $R_0$. The limit does depend
 on the phase $\phi_0$ and we present above only the absolute limit
 (${\rm max}_{\phi_0}(|1/c_0|)$). Using the analog of Eq.(\ref{rkpl}),
  this translates into
  \bea
  |F_{K^0}| < 0.0282\pm 0.0003
 \label{modf0}
  \eea
 to be compared with the limit obtained by CLEO \cite{Dobbs:2006fj}
 \bea
  |F_{K^0}| < 0.023 \ ,
 \label{modf0c}
  \eea
where the interference with the resonance has been neglected.
This analysis also implies

   \bea
  |F_{K^0}\cdot c_0| < 0.174\pm 0.009\pm 0.024 
 \label{modcapc}
  \eea
 where the first error is due to the error on $R_0$ and the
  second originates from the unknown strength of the interference
  between $A^R_{QED}$ and $A^R_{QCD}$
  (when combined the errors should added linearly).
  In the last step the hadronic amplitudes of charged
  and neutral modes are identified

   \bea
  |F_{K^0}\cdot c_0| =   |F_{K^+}\cdot c_+| \ .
   \eea

  Using $|F_{K^+}|$ as determined from Eq.(\ref{croffres}) 
 (see also Fig.\ref{fig:cphi}) one obtains experimentally allowed region 
  of  $|c_+|$ and $\phi_+$ shown in Fig.\ref{fig:cphi} as wide horizontal
 band.
  The intersection of the vertically
  dashed region (obtained from charged kaons only) and the wider horizontal
  band characterizes
  the experimentally allowed $|c_+|$ and $\phi_+$ values.
  This more conservative analysis, which does not set $F_{K^0}=0$
 and uses only the upper limit in Eq. (\ref{modf0}), thus
 leads to $|c_+|= 2.94\pm 0.99$  and no relevant constraint on
  $\phi_+$.
  
  Although the same formulae are applicable to $J/\psi$ decays
  the situation is markedly different as far as the analysis is concerned.
 At present, there are no off-resonance measurements of the cross section,
  hence only Eq.(\ref{rkpl}) (and its analog for neutral kaons) 
  can be exploited. Furthermore, the quantity $r(J/\psi) = 0.0835\pm0.0008$,
 which characterizes the contribution from the resonance--continuum
 interference to the branching ratio is significantly smaller than
 $r(\psi(2S)) = 0.663$, such that the term $r{\rm Im}c$ is significantly
  less important. Previous analyses of the $J/\psi$ branching ratios
 have systematically neglected the $r{\rm Im}c$-term
 (see e.g. \cite{Suzuki:1999nb,Rosner:1999zm,Seth:Jphi}).
 Furthermore the authors made simplifying assumptions about the 
 $\pi^+$, $K^+$ and $K^0$ form factors.
  Let us discuss these now  in detail:
   
  Isospin symmetry implies $A^R_{QCD}(K^+)=A^R_{QCD}(K^0)$ and, furthermore,
  $A^R_{QCD}(\pi^+)=0$, assumptions also used in the present analysis.
  The result for the pion form factor has been discussed above and is 
  listed in Table \ref{tab:res1}. The ratio 
  $|F_\pi|^2(\psi(2S))/|F_\pi|^2(J/\psi)\simeq 0.5$ is well compatible
  with $F_\pi \sim 1/Q^2 $ expected for the asymptotic high energy region.  
 More specific, model dependent, assumptions have been used in
 \cite{Suzuki:1999nb,Rosner:1999zm,Seth:Jphi}.

\begin{figure}[ht]
 \vspace{0.5 cm}
\begin{center}
\includegraphics[width=8.cm,height=7.cm]{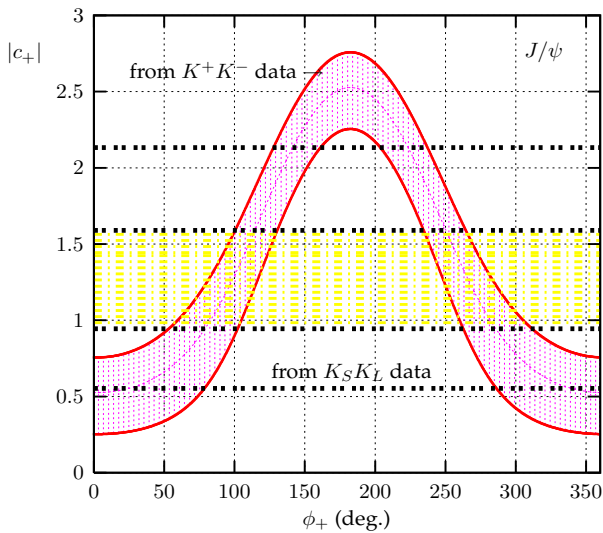}
\caption{(color online)
 The experimentally allowed region of $|c_+|$ as functions
 of $\phi_+$.
 To obtain the vertically dashed regions 
  we used $K^+K^-$ data.
 The narrow horizontal band was obtained using $A_{QED}(K^0) =0$, while 
 the wider horizontal region was obtained using neutral kaon
 data and  in addition 
 $|F_{K^+}|^2(\phi_+) = (8.03\pm 2.61)\cdot 10^{-3}$ (see text for details).
\label{fig:cphijpsi}
}
\end{center}
 \vspace{0.5 cm}
\end{figure}
 The analysis for charged and neutral kaons depends critically on the
 assumptions on the kaon form factors,
which have not yet been measured close to $J/\psi$.
  In the first step we combine
 the measured branching ratio ${\cal B}(J/\psi\to K^+K^-)$ with an
 assumption on the charged form factor $|F_{K^+}|^2=(8.0\pm2.6)\cdot
 10^{-3}$.
  The latter is obtained
 from scaling law $|F_{K^+}|$ proportional to $1/Q^2$ and using its
 measured value at the $\psi(2S)$ resonance as input. This value is
 also close to the pion form factor as expected from $SU(3)$ symmetry.
  Since the values of the hadronic
  and QED amplitudes are of comparable size the result for 
  $c_+ = A^{J/\psi}_{QCD}/A^{J/\psi}_{QED}$ depends crucially on their relative phase
 (vertically dashed region in Fig. \ref{fig:cphijpsi}). In the second
 step the branching ratio into neutral kaon is used to determine
  $A^{J/\psi}_{QCD}$. In principle ${\cal B}(J/\psi\to K^0\bar K^0)$ receives
 contributions both from $A^{J/\psi}_{QCD}$ and $A^{J/\psi}_{QED}(K^0_SK^0_L)$.
 Adopting, however, as an extreme case  $A^{J/\psi}_{QED}(K^0_SK^0_L) =0$
 the modulus of $A^{J/\psi}_{QCD}(K^0_SK^0_L) = A^{J/\psi}_{QCD}(K^+K^-)$ is fixed,
 and correspondingly $|c_+ F_{K^+}|$. Together  with the assumption
 on $|F_{K^+}|$ this leads  to $|c_+|=1.27\pm 0.32$ and the relative phase
 of $53^\circ < \phi_+ < 130^\circ$ or $234^\circ < \phi_+ < 311^\circ$.
 Although this observation is consistent with the observation of Rosner
  \cite{Rosner:1999zm} of $\phi_+$ around $90^\circ$ or $270^\circ$ 
   (see also \cite{Suzuki:1999nb}),
 our uncertainty on the phase is significantly larger,
 a consequence of the larger errors both on the branching ratio and
 the charged kaon form factor. However, even the results discussed above
 depend critically on the assumption of very small $|F_{K^0}|$.
 Although the neutral kaon form factor is indeed expected to be significantly
 smaller than $|F_{K^+}|$, a more conservative analysis  should be based on
 the upper limit on $|F_{K^0}|$ only, as derived from $|F_{K^0}|$ determined
 close to $\psi(2S)$, and the scaling law $|F_{K^0}|\sim 1/Q^2$, whence
  $|F_{K^0}|^2< 1.63\cdot 10^{-3}$ in the neighborhood of $J/\psi$.
  This implies  $0.55 < |c_+| < 2.13$ and excludes only a small region 
  between $ 161^\circ$ and $204^\circ$ for the phase $\phi_+$.
  As it is clear from this discussion, a significant improvement 
 on the charged and neutral kaon form factor measurement in the
  neighborhood of $J/\psi$ is required to disentangle hadronic
 and electromagnetic amplitudes. Let us also stress that two measurements
 combined, one several MeV, the second more then 20 MeV below the resonance,
 would allow to determine all the amplitudes up to a two-fold ambiguity.

\section{\label{sec5} Summary}
Recent experimental results for $J/\psi$ and $\Psi(2S)$ decays into
pairs of pseudoscalar mesons have been used to extract the hadronic
decay amplitude and the electromagnetic form factors at the
corresponding energies. A previously neglected interference term leads
to significant shifts of the parameters. It is demonstrated that the
previous results, observing a large relative phase between strong and
electromagnetic amplitude, depend sensitively on specific assumptions about
the neutral kaon form factor which are not tested by experiment.It is,
furthermore, shown how the combination of two cross section measurements
close to the resonance with the corresponding branching ratio would lead
to a model independent determination (up to a twofold ambiguity) of
strong amplitude, form factors and their relative phase.

\begin{acknowledgments}
Henryk Czy\.z is grateful for the support and the kind hospitality 
of the Institut f{\"u}r Theoretische Teilchenphysik
 of the Karlsruhe University. 
\end{acknowledgments}

\bibliography{biblio}

\end{document}